\documentclass[twocolumn,pra, nofootinbib, superscript address]{revtex4-2}
\usepackage[latin9]{inputenc}
\usepackage[letterpaper]{geometry}
\usepackage{color}
\usepackage{amsmath}
\usepackage{amssymb}
\usepackage{graphicx}
\usepackage[unicode=true,pdfusetitle,
 bookmarks=true,bookmarksnumbered=false,bookmarksopen=false,
 breaklinks=false,pdfborder={0 0 1},backref=false,colorlinks=true]
 {hyperref}
\hypersetup{
 pdfborderstyle=,linkcolor=blue,urlcolor=blue,citecolor=blue}

\makeatletter
\usepackage{braket}
\usepackage{upgreek}
\usepackage{lipsum}
\usepackage{enumitem}
\setlist{left= 0pt}

\makeatother

\begin{document}
\title{Coherent spatial control of wave packet dynamics on quantum lattices}
\author{Ilia Tutunnikov}
\affiliation{Department of Chemistry, Massachusetts Institute of Technology, 77
Massachusetts Avenue, Cambridge, Massachusetts 02139, USA \looseness=-1}
\author{Chern Chuang}
\affiliation{Department of Chemistry and Biochemistry, University of Nevada, 4505
S Maryland Pkwy, Las Vegas, Nevada 89154, USA \looseness=-1}
\author{Jianshu Cao}
\email{jianshu@mit.edu}

\affiliation{Department of Chemistry, Massachusetts Institute of Technology, 77
Massachusetts Avenue, Cambridge, Massachusetts 02139, USA \looseness=-1}
\begin{abstract}
Quantum lattices are pivotal in the burgeoning fields of quantum materials
and information science. Rapid developments in microscopy and quantum
engineering allow for preparing and monitoring wave-packet dynamics
on quantum lattices with increasing spatial and temporal resolution.
Motivated by these emerging research interests, we present an analytical
study of wave packet diffusivity and diffusion length on tight-binding
quantum lattices subject to stochastic noise. Our analysis points
to the crucial role of spatial coherence and predicts a set of novel
phenomena: noise can enhance the transient diffusivity and diffusion
length of sufficiently extended initial states; A smooth Gaussian
initial state spreads slower than a localized initial state; A standing
or traveling initial state with large momentum spreads faster than
a localized initial state and exhibits a noise-induced peak in the
transient diffusivity; The change in the time-dependent diffusivity
and diffusion length relative to a localized initial state follows
a universal dependence on the Gaussian width. These theoretical predictions
and the underlying mechanism of spatial coherence suggest the possibility
of controlling the wave packet dynamics on quantum lattices by spatial
manipulations, which will have implications for materials science
and quantum technologies.
\end{abstract}
\maketitle
The rapid advancement in quantum materials and quantum information
science has generated renewed interest in quantum lattices \citep{Goringe1997}
of two archetypes: molecular crystals and quantum engineering platforms.
Molecular crystals \citep{KuhnBook2011} encompass a broad range of
organic and inorganic compounds and have demonstrated remarkable potential
in charge \citep{Karl1999,Podzorov2004,Sakanoue2010} and excited
state energy transport \citep{Silbey1976,Dykstra2009,Moix2013,Bardeen2014,Lee2015,Chuang2016,Doria2018,Maly2020,Popp2021,Chuang2021,Blach2022}
in terms of tunability and cost-effectiveness. These materials, often
characterized by their well-ordered structures, are an ideal platform
for quantum coherent control studies. On the other hand, quantum platforms,
including optical lattices \citep{Dahan1996,Preiss2015} and superconducting
circuits \citep{Wang2020,Gong2021,Karamlou2022}, offer an entirely
different approach to engineering systems for quantum information
applications. In particular, such systems can emulate various types
of quantum random walks \citep{Childs2002,Kempe2003,Mulken2005}.
Exploring these two lattice classes extends our understanding of fundamental
quantum phenomena and opens new avenues for advancing quantum technologies.

This work delves into the transient diffusivity and diffusion length
of spatially extended states on quantum lattices. The confluence of
cutting-edge quantum technology and advanced microscopy techniques
allows the preparation and observation of quantum phenomena at unprecedented
spatial and temporal resolutions \citep{Akselrod2014,Zhu2019,Ginsberg2020,Delor2020}.
Thus, we shift the focus from traditional coherent temporal control
\citep{Tannor1985,Brumer1986,Cao1998,RiceBook2000,Chakrabarti2007,TannorBook2007,ShapiroBrumer2011}
to spatial coherent control. It is a step toward studying quantum
phenomena in solid rather than gaseous phases, including quantum materials.

In this paper, we first introduce various extended wave packets in
a tight-binding lattice and then present analytical expressions for
their diffusivity and diffusion length in closed and open systems.
Our analysis reveals an array of intriguing quantum phenomena, including
the noise-induced speed-up and a universal scaling of transient diffusivity
and diffusion length with the wave packet width. These findings underscore
the crucial role of spatial coherence in the quantum evolution.\\

\noindent \textbf{System Definitions and General Formulas.} We examine
the dynamics of wave packets on a one dimensional (1D) homogeneous
tight-binding lattice, which is an infinite chain of sites with inter-site
coupling strength $J_{m,n}$ between sites $m$ and $n$. The Hamiltonian
describing the couplings is
\begin{equation}
\hat{H}=\sum_{m\neq n}J_{m,n}\ket{m}\bra{n},\label{eq:H-tight-binding}
\end{equation}
where $\ket{m}$ and $\ket{n}$ denote the localized states at sites
$m$ and $n$. The physical position of the $n$-th site (assuming
an equally-spaced lattice) is given by $x_{n}=an$, where $a$ is
the lattice constant. All lengths in this study are normalized to
$a$. We set $\hbar=1$, and measure time in units of inverse energy.

The primary observable in this study is the time-dependent diffusivity,
$D(t)$ defined as
\begin{equation}
2D(t)=\frac{d\braket{n^{2}(t)}}{dt}-\frac{d\braket{n(t)}^{2}}{dt},\label{eq:D-definition}
\end{equation}
where $\langle n^{k}(t)\rangle=\sum_{n}n^{k}\rho_{n,n}(t),$ and $\rho_{n,n}(t)$
are the diagonal elements of the density matrix. For analytical tractability,
we restrict our analysis to nearest-neighbor coupling, $J_{m,n}=J\delta_{m,n\pm1}$.
Under this assumption, $\rho_{n,n}(t)$ satisfy the equation $\dot{\rho}_{n,n}(t)=2J\mathrm{Im}[\rho_{n-1,n}(t)+\rho_{n+1,n}(t)]$,
where we used $\dot{\rho}=-i[H,\rho]=-i(H\rho-\rho H)$, and $H$
is the matrix representation of $\hat{H}$ in the basis of localized
states. Spatial coherences and their first moments are defined as
($l\geq0$)
\begin{align}
\braket{\rho(t)}_{l}=\sum_{n}\rho_{n,n+l}(t),\quad\braket{n(t)}_{l} & =\sum_{n}n\rho_{n,n+l}(t).\label{eq:<n>_l-<=00005Crho>_l-defs}
\end{align}
In this notation,
\begin{align}
\frac{d\langle n(t)\rangle}{dt} & =\sum_{n}n\dot{\rho}_{n,n}(t)=2J\mathrm{Im}[\braket{\rho(t)}_{1}],\\
\frac{d\langle n^{2}(t)\rangle}{dt} & =\sum_{n}n^{2}\dot{\rho}_{n,n}(t)=4J\mathrm{Im}[\braket{n(t)}_{1}]+\frac{d\langle n(t)\rangle}{dt},
\end{align}
where $J\mathrm{Im}[\braket{\rho(t)}_{1}]$ can be interpreted as
population flux, and $J\mathrm{Im}[\braket{n(t)}_{1}]$ is the displacement-weighted
flux.

When the wave packet center of mass is stationary ($d\langle n(t)\rangle/dt=0$),
the diffusivity simplifies to
\begin{equation}
D(t)=\frac{1}{2}\frac{d\braket{n^{2}(t)}}{dt}=2J\mathrm{Im}[\braket{n(t)}_{1}],\label{eq:D(t)=00003D2JIm=00005B<n(t)>_2=00005D}
\end{equation}
which holds for nearest neighbor coupling even in the presence of
noise (considered below). $\mathrm{Im}[\braket{n(t)}_{1}]$ satisfies
the following equation
\begin{equation}
\frac{d\mathrm{Im}[\braket{n(t)}_{1}]}{dt}=\mathcal{S}[\braket{n(t)}_{1}]-J\mathrm{Re}[\braket{\rho(t)}_{2}],\label{eq:dn_1/dt-S-operator}
\end{equation}
where $\mathrm{Re}[\braket{\rho(t)}_{2}]$ is a measure of spatial
coherence. The function $\mathcal{S}[\braket{n(t)}_{1}]$ varies depending
on whether the system is isolated or subject to noise; its explicit
forms will be specified later.

We analyze the dynamics of two types of initial states. The first
type is a standing Gaussian wave packet with a stationary center of
mass, described by
\begin{align}
\psi_{n}(0)= & \frac{\sqrt{2}\cos(kn)}{\sqrt{w\sqrt{\pi}[1+e^{-k^{2}w^{2}}]}}\exp\left[-\frac{n^{2}}{2w^{2}}\right],\label{eq:in-state-standing}
\end{align}
where $k$ is the wave number, and $w>0$ is the initial width. The
standard Gaussian is obtained by setting $k=0$. The second initial
state is the traveling Gaussian
\begin{equation}
\psi_{n}(0)=\frac{1}{\sqrt{w\sqrt{\pi}}}\exp\left[-\frac{n^{2}}{2w^{2}}+ipn\right],\label{eq:in-state-traveling}
\end{equation}
where $p$ is the initial momentum of the wave packet. The traveling
Gaussian state describes wave-like propagation and can be relevant
for cavity polaritons \citep{EbbesenReview2021,Engelhardt2022,Cao2022,Engelhardt2023}.
Here, we restrict the magnitude of $k$ and $p$, $0\leq|k|,|p|\leq\pi/2$.
Extending the values beyond this range yields periodic repetitions
of the results.\\

\noindent \textbf{Isolated System.} This section focuses on analyzing
diffusivity in an isolated system, while the effects of noise will
be explored in subsequent sections. In an isolated system, $\mathcal{S}[\braket{n(t)}_{1}]=J$,
and the spatial coherences are conserved, $\braket{\rho(t)}_{l}=\braket{\rho(0)}_{l}$,
$l\geq0$. Consequently, for wave packets with a stationary center
of mass, the diffusivity is given by
\begin{equation}
D(t)=D_{\delta}(t)-2J^{2}\mathrm{Re}[\braket{\rho(0)}_{2}]t,\label{eq:D(t)-stationary-CM}
\end{equation}
where $D_{\delta}(t)=2J^{2}t$ is the well-known diffusivity of Kronecker
delta initial state. As indicated by Eq. \eqref{eq:D(t)-stationary-CM},
a positive value of spatial coherence $\mathrm{Re}[\braket{\rho(0)}_{2}]>0$
leads to suppressed diffusivity relative to $D_{\delta}$, and vice
versa for $\mathrm{Re}[\braket{\rho(0)}_{2}]<0$. For wave packets
with a non-stationary center of mass, the expressions for diffusivity
and the center of mass are as follows
\begin{align}
D(t) & =D_{\delta}(t)-2J^{2}(\mathrm{Re}[\braket{\rho(0)}_{2}]+2\mathrm{Im}[\braket{\rho(0)}_{1}]{}^{2})t\nonumber \\
 & +J(\mathrm{Im}[\braket{\rho(0)}_{1}]+2\mathrm{Im}[\braket{n(0)}_{1}]),\label{eq:D-formula}\\
\braket{n(t)} & =2J\mathrm{Im}[\braket{\rho(0)}_{1}]t.\label{eq:<n>-formula}
\end{align}
Here, without loss of generality, we assume $\braket{n(0)}=0$. For
detailed derivation and explicit expressions for $\braket{\rho(0)}_{1,2}$
and $\braket{n(0)}_{1}$ for the considered wave packets, see SI.
Note that throughout this work, the sums $\braket{\rho(0)}_{1,2}$
and $\braket{n(0)}_{1}$ were approximated by integrals. This approximation
has a negligible effect for initial states with a width equal to or
greater than a single lattice constant $a$.

\begin{figure}[t]
\begin{centering}
\includegraphics{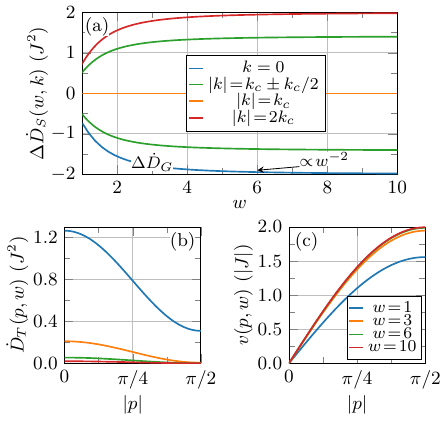}
\par\end{centering}
\centering{}\caption{Isolated system. (a) Derivative of relative diffusivity of a standing
Gaussian initial state, see Eq. \eqref{eq:dtDeltaD_S}. (b) Derivative
of diffusivity of a traveling Gaussian initial state, see Eq. \eqref{eq:D_T}.
(c) Center of mass velocity, see Eq. \eqref{eq:<n>}.}
\label{fig:FIG1}
\end{figure}

\textbf{\emph{Isolated system I -- Gaussian initial state.}}--For
the Gaussian initial state, $\braket{\rho(0)}_{2}=\exp(-1/w^{2})$,
such that
\begin{equation}
D_{G}(t)=2J^{2}(1-e^{-1/w^{2}})t,\quad D_{G}(t)\approx\frac{2J^{2}}{w^{2}}t,\;w\gg1.\label{eq:D_G}
\end{equation}
The larger the initial width $w$, the higher the spatial coherence
and the lower the diffusivity. Note that the relative diffusivity
is strictly negative $\Delta\dot{D}_{G}\equiv\dot{D}_{G}-\dot{D}_{\delta}=-2J^{2}\exp(-1/w^{2})<0$,
and for $w\gg1$, $\Delta\dot{D}_{G}$ scales as $w^{-2}$, as illustrated
in Fig. \ref{fig:FIG1}(a).

Qualitatively, the decay of diffusivity with increasing $w$ can be
understood in terms of the uncertainty principle. The squared expansion
coefficient of the Gaussian state in terms of Bloch states, $\propto\exp(in\nu)$
is $|c_{\nu}|^{2}\propto\exp(-w^{2}\nu^{2})$, $\nu\in[-\pi,\pi]$.
$|c_{\nu}|^{2}$ shrinks towards $\nu=0$ with increasing $w$, while
the dispersion relation reads $E_{\nu}=2J\cos(\nu)\approx2J(1-\nu^{2}/2)$.
Consequently, an increase in $w$ results in reduced energy dispersion
among the wave packet's components, leading to slower spatial spreading.
Formally, in the limit $w\rightarrow\infty$, the wave packet becomes
a soliton---a wave packet that retains its shape. However, this picture
is complicated in open systems by the presence of noise, a topic that
will be explored in subsequent sections.

Quantitatively, the diffusivity is given by the Green-Kubo formula
\begin{equation}
D_{G}(t)=\int_{0}^{t}\braket{v_{g}(t')v_{g}(0)}\,dt'=t\int_{-\pi}^{\pi}v_{g}^{2}|c_{\nu}|^{2}\,d\nu,\label{eq:D_G-GK}
\end{equation}
where $v_{g}(t')\equiv v_{g}=d_{\nu}E_{\nu}=-2J\sin(\nu)$ is the
time-independent group velocity of state $\nu$, and $\braket{\cdot}$
denotes the weighted average over Bloch states. A sharper distribution
$|c_{\nu}|^{2}$ results in a smaller $\dot{D}_{G}$. The integral
in Eq. \eqref{eq:D_G-GK} can be explicitly evaluated, thereby reproducing
Eq. \eqref{eq:D_G} (see SI for details).

\begin{figure*}
\begin{centering}
\includegraphics{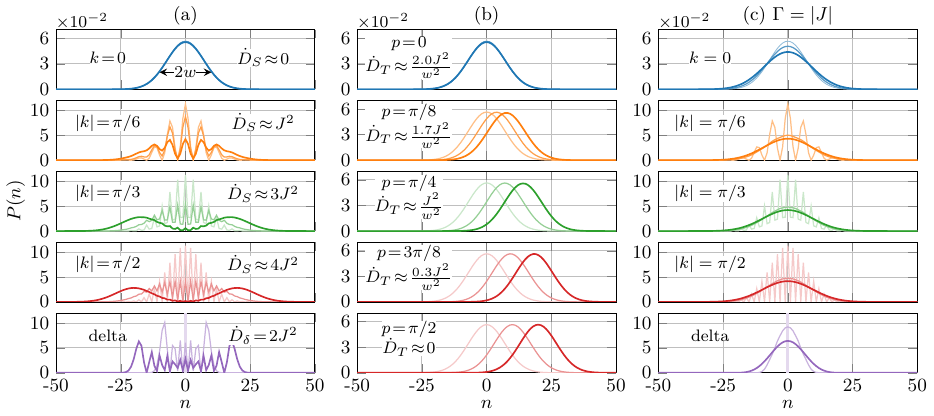}
\par\end{centering}
\centering{}\caption{Population distributions. (a) Isolated system, standing Gaussian.
(b) Isolated system, traveling Gaussian. Here $J<0$. (c) Open system
(with noise), standing Gaussian. Thin, intermediate, and thick lines
correspond to $t=0,\,5/|J|,\,10/|J|$. $w=10$.}
\label{fig:FIG2}
\end{figure*}

\textbf{\emph{Isolated system II -- Standing Gaussian initial state.}}--We
now focus on the general case described by Eq. \eqref{eq:in-state-standing},
where $k\neq0$. Substituting $\braket{\rho(0)}_{2}$ for the standing
Gaussian initial state into Eq. \eqref{eq:D(t)-stationary-CM} yields
\begin{align}
D_{S}(t) & =D_{\delta}-2J^{2}e^{-1/w^{2}}\frac{e^{k^{2}w^{2}}\cos(2k)+1}{e^{k^{2}w^{2}}+1}t\nonumber \\
 & \approx2J^{2}\left[1-e^{-1/w^{2}}\cos(2k)\right]t;\;w\gg1,\label{eq:D_S-w>>1}
\end{align}
where $\braket{\rho(0)}_{2}\approx\exp(-1/w^{2})\cos(2k)$. Figure
\ref{fig:FIG1}(a) shows the derivative of relative diffusivity
\begin{equation}
\Delta\dot{D}_{S}\equiv\dot{D}_{S}-\dot{D}_{\delta}\approx-2J^{2}e^{-1/w^{2}}\cos(2k).\label{eq:dtDeltaD_S}
\end{equation}
The expression implies the critical wave number $|k|=k_{c}\equiv\pi/4$,
at which $\Delta D_{S}$ vanish. For $|k|>k_{c}$, $D_{S}(t)$ is
enhanced relative to $D_{\delta}(t)$, and vice versa for $|k|<k_{c}$.

Figure \ref{fig:FIG2}(a) depicts the population distributions for
a standing Gaussian initial state with $w=10$, evaluated at different
times $t=0,\,5/|J|,\,10/|J|$, and for various values of $|k|$. The
numerical results confirms the prediction of Eqs. \eqref{eq:D_S-w>>1}
and \eqref{eq:dtDeltaD_S}. For the Gaussian state ($k=0$), there
is no visible change on the considered time scale. For $|k|>0$, the
wave function $\psi_{n}(0)\propto[\exp(ikn)+\exp(-ikn)]\exp[-n^{2}/(2w^{2})]$
gives rise to two wave packets that move in opposite directions with
momenta $\pm k$. Initially, the overlapping of the two wave packets
results in an interference pattern; however, as they separate, this
interference diminishes. Similar to the case of a single Gaussian,
an increase in the initial width $w$ leads to reduced energy dispersion
and a slower rate of expansion relative to each Gaussian's center
of mass. Consequently, when $w\gg1$, the dominant contribution to
the diffusivity comes from the relative separation of the two wave
packets.

\textbf{\emph{Isolated system III -- Traveling Gaussian initial state.--}}We
now turn to the traveling Gaussian initial state, as described by
Eq. \eqref{eq:in-state-traveling}, which is similar to one of the
two Gaussians in the standing Gaussian state. Substituting $\braket{\rho(0)}_{1,2}$
and $\braket{n(0)}_{1}$ into Eqs. \eqref{eq:D-formula} and \eqref{eq:<n>-formula},
yields the diffusivity $D_{T}(t)$ and the center of mass position
$\braket{n(t)}$,
\begin{align}
D_{T}(t) & \!=\!2J^{2}[1\!-\!e^{-1/w^{2}}\cos(2p)\!-\!2e^{-1/(2w^{2})}\sin^{2}(p)]t,\label{eq:D_T}\\
\braket{n(t)} & =-2Je^{-1/(4w^{2})}\sin(p)t.\label{eq:<n>}
\end{align}
Here, the range of initial momentum magnitude is $0\leq|p|\leq\pi/2$.
Figure \ref{fig:FIG1}(b) illustrates that an increase in either $|p|$
or $w$ leads to suppression of $\dot{D}_{T}$. When $w\gg1$,
\begin{equation}
\dot{D}_{T}\approx\frac{J^{2}}{w^{2}}[1+\cos(2p)]\leq\dot{D}_{G},\label{eq:D_T-approx.}
\end{equation}
which is consistent with the large $w$ limit of Eq. \eqref{eq:D_G},
when $|p|\ll1$. The motion becomes soliton-like for $|p|=\pi/2$,
i.e., the wave packet retains its shape (refer to SI for a qualitative
discussion). Figure \ref{fig:FIG1}(c) shows that the speed $v\equiv|d_{t}\braket{n}|$
is constant over time and increases with both the initial width, $w$
and $|p|$. Figure \ref{fig:FIG2}(b) shows population distributions
for the traveling Gaussian with $w=10$ and several values of $p$.
There is no visible spreading of the moving wave packet, as the diffusivity
is suppressed by the factor of $w^{-2}$.

The diffusivities of standing and traveling Gaussian initial states
are related by $D_{S}(t)=D_{T}(t)+d_{t}\braket{n(t)}^{2}/2$, where
$D_{S}(t)$ is the diffusivity of the standing Gaussian with $k$
substituted by $p$ {[}see Eq. \eqref{eq:D_S-w>>1}{]}. This relationship
is consistent with the observations illustrated in Fig. \ref{fig:FIG2}.
In particular, the enhanced diffusivity in the standing Gaussian state
is mainly attributed to the relative motion between the two Gaussians,
as indicated by the second term, rather than the slower dispersion
around each Gaussian's center of mass, which is represented by the
first term scaling as $w^{-2}$.\\

\noindent \textbf{Dynamics in the Presence of Noise: HSR Model.} In
this section, we explore the impact of noise on diffusivity within
the framework of the Haken-Strobl-Reineker (HSR) model, a well-established
approach for the high-temperature regime \citep{Madhukar1977,ExcitonDynamicsBook1982,Amir2009}.
The model describes site energy fluctuations as white noise. At sufficiently
high temperatures, thermal fluctuations dominate over static disorder,
allowing us to consider a homogeneous chain. The role of static disorder
will be explored in future work \citep{Tutunnukov2023InProgress},
particularly in the context of Anderson localization.

Within the HSR model, the density matrix evolves according to the
Liouvillian equation \citep{ExcitonDynamicsBook1982}
\begin{align}
\dot{\rho} & =-i[H,\rho]-\frac{\Gamma}{2}\sum_{n=-\infty}^{\infty}[V_{n},[V_{n},\rho]],\label{eq:HSR-model}
\end{align}
where the matrix elements of $V_{n}$ are $(V_{n})_{j,k}=\delta_{j,n}\delta_{k,n}$,
and $\Gamma$ is the dephasing rate. Another way of writing Eq. \eqref{eq:HSR-model}
is $\begin{aligned}\dot{\rho} & =-i[H,\rho]-\Gamma\rho_{H}\end{aligned}
$ with $(\rho_{H})_{m,n}=(1-\delta_{m,n})\rho_{m,n}$. This form emphasizes
the suppression of the off-diagonal elements of the density matrix
(coherences), manifesting in exponential decay of spatial coherences,
$\braket{\rho(t)}_{l}=\braket{\rho(0)}_{l}\exp(-\Gamma t)$, $l\geq1$
(the proof is outlined in SI). In the HSR model, $\mathcal{S}(\braket{n(t)}_{1})=J-\Gamma\mathrm{Im}[\braket{n(t)}_{1}]$
{[}see Eq. \eqref{eq:dn_1/dt-S-operator}{]}, such that
\begin{equation}
\frac{d\mathrm{Im}[\braket{n(t)}_{1}]}{dt}=J-\Gamma\mathrm{Im}[\braket{n(t)}_{1}]-J\mathrm{Re}[\braket{\rho(t)}_{2}].\label{eq:d<n_1(t)>/dt-HSR}
\end{equation}
This is a closed-form equation, amenable to analytical solutions.
For real initial states with stationary center of mass, we obtain
the generalized form of Eq. \eqref{eq:D(t)-stationary-CM}
\begin{align}
D_{S}(t) & =D_{\delta}(t)-2J^{2}\mathrm{Re}[\braket{\rho(t)}_{2}]t\nonumber \\
 & =D_{\delta}(t)-2J^{2}\mathrm{Re}[\braket{\rho(0)}_{2}]e^{-\Gamma t}t.\label{eq:D_S-HSR-general}
\end{align}
In the presence of noise, $D_{\delta}$ is given by \citep{Moix2013}
\begin{eqnarray}
D_{\delta}(t) & = & 2J^{2}\frac{1-e^{-\Gamma t}}{\Gamma}.\label{eq:D-Gamma-delta-state}
\end{eqnarray}
On short time scale ($\Gamma t\ll1$), the effect of dephasing is
negligible, and one recovers the ballistic behavior i.e., $D_{\delta}(t)\propto t$.
The coherence is completely lost in the long time limit, leading to
a constant diffusivity, $D_{\delta}(t\rightarrow\infty)=2J^{2}/\Gamma$.

\textbf{\emph{Open system I -- Gaussian initial state.}}--For the
Gaussian initial state, $\braket{\rho(0)}_{2}=\exp(-1/w^{2})$, and
substitution into Eq. \eqref{eq:D_S-HSR-general} yields
\begin{align}
D_{G}(t) & =2J^{2}\left[\frac{1-e^{-\Gamma t}}{\Gamma}-e^{-1/w^{2}}e^{-\Gamma t}t\right].\label{eq:D_G-HSR}
\end{align}

\begin{figure}[t]
\begin{centering}
\includegraphics{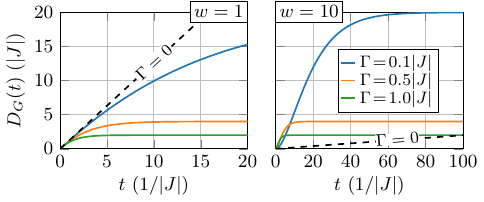}
\par\end{centering}
\centering{}\caption{Diffusivity in the presence of noise for Gaussian initial state of
width $w$, see Eq. \eqref{eq:D_G-HSR}.}
\label{fig:FIG3}
\end{figure}

\begin{itemize}
\item On the short time scale ($\Gamma t\ll1$), the diffusivity simplifies
to that given by Eq. \eqref{eq:D_G}. 
\item In the long-time limit, the diffusivity reaches a steady-state value
of $D_{G}(t\rightarrow\infty)=2J^{2}/\Gamma$, independent of the
initial width $w$. 
\item As in the isolated system $D_{G}(t)$ remain smaller than $D_{\delta}(t)$,
\begin{equation}
\Delta D_{G}(t)\equiv D_{G}(t)-D_{\delta}(t)=-2J^{2}e^{-1/w^{2}}e^{-\Gamma t}t\leq0.\label{eq:DeltaD_G-HSR}
\end{equation}
This difference also depends on $w$ through factor $\exp(-1/w^{2})$.
\end{itemize}
Figure \ref{fig:FIG3} shows $D_{G}(t)$ for several dephasing rates
$\Gamma$. For narrow initial states, such as $w=1$, noise acts to
suppress diffusivity. Conversely, for wider states like $w=10$, diffusivity
experiences a transient enhancement. This behavior is in striking
contrast to the diffusivity suppression with increasing $w$ in the
isolated system. The \emph{noise-induced enhancement of transient
diffusivity} in relatively wide initial states is one of the main
results of this letter.

The critical width $w_{c}$ for observing an enhancement in diffusivity
can be determined by examining the sign of $\Delta_{\Gamma}D_{G}\equiv D_{G}(t,\Gamma\neq0)-D_{G}(t,\Gamma=0)$.
For $\Gamma t\ll1$, the expression simplifies to $\Delta_{\Gamma}D_{G}\approx\Gamma J^{2}t^{2}[2\exp(-1/w^{2})-1]$,
such that
\begin{equation}
w_{c}=\frac{1}{\sqrt{\ln(2)}}\approx1.2.\label{eq:Gaussain-w_c}
\end{equation}

\textbf{\emph{Open system II -- Standing Gaussian initial state.}}--For
the standing Gaussian initial state, $\braket{\rho(t)}_{2}=\braket{\rho(0)}_{2}\exp(-\Gamma t)\approx\exp(-1/w^{2})\cos(2k)\exp(-\Gamma t)$,
such that
\begin{align}
D_{S}(t) & =D_{\delta}-2J^{2}e^{-1/w^{2}}\frac{e^{k^{2}w^{2}}\cos(2k)+1}{e^{k^{2}w^{2}}+1}e^{-\Gamma t}t\nonumber \\
\approx2J^{2} & \left[\frac{1-e^{-\Gamma t}}{\Gamma}-e^{-1/w^{2}}\cos(2k)e^{-\Gamma t}t\right],\;w\gg1.\label{eq:D_S-HSR}
\end{align}
The critical width $w_{c}$ now depends on the wave number $k$
\begin{equation}
w_{c}\approx\frac{1}{\sqrt{\ln[2\cos(2k)]}}.\label{eq:Standing-w_c}
\end{equation}
Noise-induced enhancement is achievable only when $|k|<\pi/6$.

\begin{figure}[h]
\begin{centering}
\includegraphics{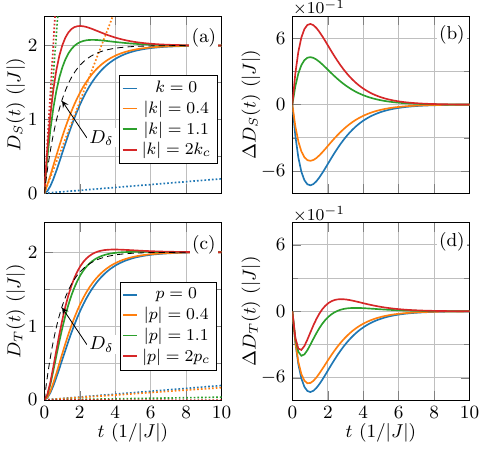}
\par\end{centering}
\centering{}\caption{Diffusivity and relative diffusivity in the presence of noise. (a,
b) Standing Gaussian initial state, see Eqs. \eqref{eq:D_S-HSR} and
\eqref{eq:DeltaD_S-HSR}. (c, d) Traveling Gaussian initial state,
see Eq. \eqref{eq:D_T-HSR}. $w=10$, $\Gamma=|J|$. The corresponding
diffusivities in the isolated system ($\Gamma=0$) are shown by dotted
lines in (a) and (c).}
\label{fig:FIG4}
\end{figure}

Fig. \ref{fig:FIG4}(a) shows $D_{S}(t)$ for a range of $|k|$ values.
On the short time scale ($\Gamma t\ll1$), the diffusivity is consistent
with Eq. \eqref{eq:D_S-w>>1}. On the long time scale, regardless
of the initial state, $D_{S}\rightarrow2J^{2}/\Gamma$. For $|k|\leq\pi/6$
($|k|=0,0.4$), there is transient diffusivity enhancement where the
solid lines are above the corresponding dotted lines. 

For $|k|>k_{c}\equiv\pi/4$, $D_{S}(t)$ exhibits a peak at $t_{p}\approx[1-\exp(1/w^{2})\sec(2k)]/\Gamma$.
This peak is most pronounced when $|k|=2k_{c}$, occurring at $t_{p}=[1+\exp(1/w^{2})]/\Gamma$.
Wave number $2k_{c}$ corresponds to a wave length of four lattice
spacings, manifesting in destructive interference in spatial coherence.
The peak signifies the transition from ballistic wave packet expansion
to the steady-state diffusion. For $|k|>k_{c}$, the standing Gaussian
initially expands more rapidly than the Kronecker delta. Dephasing,
however, takes over eventually, lowering the diffusivity towards the
steady-state value. The relative diffusivity is given by
\begin{equation}
\Delta D_{S}(t)\equiv D_{S}(t)-D_{\delta}(t)\!=\!-2J^{2}e^{-1/w^{2}}\cos(2k)e^{-\Gamma t}t.\label{eq:DeltaD_S-HSR}
\end{equation}
It depends on $w$ through the factor $\exp(-1/w^{2})$, analogous
to the Gaussian case described by Eq. \eqref{eq:DeltaD_G-HSR}. Figure
\ref{fig:FIG4}(b) shows $\Delta D_{S}(t)$ corresponding to panel
(a). Once again, we see that for $|k|>k_{c}$ the diffusivity is enhanced
relative to $D_{\delta}$, and vice versa for $|k|<k_{c}$. 

The dynamics of population distribution in the presence of noise,
as illustrated in Fig. \ref{fig:FIG2}(c), also differs from the isolated
system. For dephasing rate value of $\Gamma=|J|$, noise quickly annihilates
spatial coherence and by the time $t=5/|J|$, the distributions for
all $|k|$ values smooth out.

\textbf{\emph{Open system III -- Traveling Gaussian initial state.}}--For
the traveling Gaussian initial state, the diffusivity and the center
of mass position expressions are (for details, see SI)
\begin{align}
D_{T}(t) & =2J^{2}\frac{1-e^{-\Gamma t}}{\Gamma}-2J^{2}e^{-1/w^{2}}\cos(2p)e^{-\Gamma t}t\nonumber \\
- & \frac{4J^{2}}{\Gamma}e^{-1/(2w^{2})}\sin^{2}(p)(1-e^{-\Gamma t})e^{-\Gamma t},\label{eq:D_T-HSR}\\
\braket{n(t)} & =-\frac{2J}{\Gamma}e^{-1/(4w^{2})}\sin(p)(1-e^{-\Gamma t}).\label{eq:<n>-HSR}
\end{align}

On the short time scale ($\Gamma t\ll1$), the center of mass moves
with a constant velocity, consistent with the isolated system, see
Eq. \eqref{eq:<n>}. However, the center of mass motion halts due
to noise once $\braket{n(t\rightarrow\infty)}=(-2J/\Gamma)\exp[-1/(4w^{2})]\sin(p)$
is reached. A larger initial momentum magnitude $|p|$ and width $w$
allow the center of mass to travel a greater distance. While in the
isolated system the traveling Gaussian shows a soliton-like motion
for $p=\pi/2$, the center of mass motion slows down in the presence
of noise. Moreover, the diffusivity tends to $p$-independent steady-state
value.

Figure \ref{fig:FIG4}(c) shows $D_{T}(t)$ for several $|p|$ values.
The critical width for the traveling Gaussian depends on $|p|$, and
varies between $w_{c}(p=0)=[\ln(2)]^{-1/2}\approx1.2$ {[}see Eq.
\eqref{eq:Gaussain-w_c}{]} and $w_{c}(|p|=\pi/2)=[2\ln(3+\sqrt{7})]^{-1/2}\approx0.54$.
Thus, $w=10$ in Fig. \ref{fig:FIG4} is well above the critical width
for all $|p|$ values. The diffusivity does not necessarily approach
the steady-state limit in a monotonic fashion. For $|p|>p_{c}\equiv\pi/4$,
$D_{T}(t)$ has a peak. The peak is most pronounced for $p=2p_{c}$
with $t_{p}\approx3.92/\Gamma$. Figure \ref{fig:FIG4}(d) show $\Delta D_{T}(t)$
corresponding to panel (c). Relative to diffusivity of a delta function
initial state, the diffusivity may be transiently enhanced for $|p|>p_{c}$
or suppressed for $|p|<p_{c}$. \\

\noindent \textbf{Diffusion Length.} Another experimental observable
that measures the expansion of finite width initial states is the
diffusion length, $L$ \citep{Lunt2009,Lin2014}. To define $L$,
consider the relative mean square displacement (MSD), $R^{2}(t)=\braket{n^{2}(t)}-\braket{n^{2}(0)}$,
which is related to the diffusivity by $R^{2}(t)=\int_{0}^{t}2D(t')\,dt'.$
In realistic scenarios, an exciton has a finite lifetime due to both
radiative and non-radiative decay mechanisms. For simplicity, we model
the exciton lifetime $t$ as a Poisson process with an exponential
distribution $P(t)=\tau^{-1}\exp(-t/\tau),$ where $\tau\geq0$ is
the average lifetime. $L^{2}$ is defined as the average of $R^{2}(t)$,
\begin{align}
L^{2} & =\int_{0}^{\infty}R^{2}(t)P(t)\,dt.\label{eq:L^2-def.}
\end{align}

In the presence of noise, $L^{2}$ of a Gaussian initial state is
given by 
\begin{equation}
L^{2}=2\int_{0}^{\infty}\!\!\int_{0}^{t}\left[D_{\delta}(t')\!-\!2J^{2}\mathrm{Re}[\braket{\rho(t')}_{2}]t'\right]P(t)\,dt'dt,
\end{equation}
where we used Eq. \eqref{eq:D_S-HSR-general}, with $\braket{\rho(t)}_{2}=\braket{\rho(0)}_{2}\exp(-\Gamma t)=\exp(-1/w^{2})\exp(-\Gamma t)$,
and $D_{\delta}(t)$ from Eq. \eqref{eq:D-Gamma-delta-state}. Explicitly,
\begin{align}
L^{2} & =4J^{2}\tau^{2}\frac{1+\Gamma\tau-e^{-1/w^{2}}}{(\Gamma\tau+1)^{2}}.\label{eq:L^2-Gaussian}
\end{align}

\begin{figure}[h]
\begin{centering}
\includegraphics{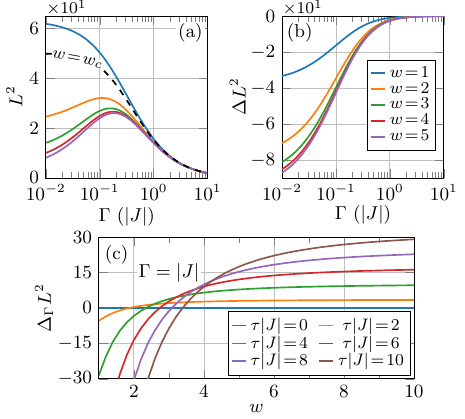}
\par\end{centering}
\centering{}\caption{(a) Squared diffusion length of a Gaussian initial state of width
$w$, see Eq. \eqref{eq:L^2-Gaussian}. $w_{c}=[\ln(2)]^{-1/2}$,
see Eq. \eqref{eq:Gaussain-w_c}. (b) Relative diffusion length, $\Delta L^{2}$
{[}see Eq. \eqref{eq:DeltaL^2}{]}. In (a) and (b), $\tau|J|=5$.
(c) Relative diffusion length, $\Delta_{\Gamma}L^{2}$.}
\label{fig:FIG5}
\end{figure}

Figure \ref{fig:FIG5}(a) shows $L^{2}$ as functions of $\Gamma$
for an average lifetime $\tau=5/|J|$ and several initial widths $w$.
When $\Gamma/|J|\ll1$, the diffusivity is suppressed as $w$ increases
due to growing spatial coherence $\braket{\rho(0)}_{2}=\exp(-1/w^{2})$.
However, for sufficiently wide initial states {[}$w>w_{c}$, as per
Eq. \eqref{eq:Gaussain-w_c}{]}, noise can transiently enhance diffusivity,
as shown in Fig. \ref{fig:FIG3}, leading to non-monotonic behavior
in $L^{2}$. When $w>w_{c}$, $L^{2}$ has a maximum at $\Gamma_{m}=[2\exp(1/w^{2})-1]/\tau$.
In the limit $\Gamma\tau\gg1$, $L^{2}$ becomes independent of $w$,
and the diffusive transport, $L^{2}\propto\tau/\Gamma$ is recovered. 

To compare the diffusion lengths of a Gaussian and Kronecker delta
initial states, we consider the relative diffusion length
\begin{equation}
\Delta L^{2}\equiv L^{2}-L_{\delta}^{2}=-4J^{2}e^{-1/w^{2}}\frac{\tau^{2}}{(\Gamma\tau+1)^{2}}.\label{eq:DeltaL^2}
\end{equation}
$\Delta L^{2}$ is proportional to the spatial coherence $\braket{\rho(0)}_{2}=\exp(-1/w^{2})$,
consistent with $\Delta D_{G}$ in Eq. \eqref{eq:DeltaD_G-HSR}. Figure
\ref{fig:FIG5}(b) shows $\Delta L^{2}$ as a function of $\Gamma$
for a fixed average lifetime $\tau=5/|J|$ and several initial widths
$w$. For negligible dephasing rate, $\Gamma\ll|J|$, we recover the
isolated system limit, where $\braket{\rho(0)}_{2}$ suppresses diffusivity
with increasing $w$. In this regime, the diffusion length of a Gaussian
is significantly lower compared to the Kronecker delta. For intermediate
values of $\Gamma$, noise allows the wave packet to expand more efficiently,
approaching the size reached by the Kronecker delta initial state
for the same average exciton lifetime. When $\Gamma\gg|J|$, the differences
between various initial states become negligible.

Next, we examine the gain in diffusion length due to noise using $\Delta_{\Gamma}L^{2}\equiv L^{2}(\Gamma)-L^{2}(0)$,
\begin{equation}
\Delta_{\Gamma}L^{2}=4\Gamma J^{2}\tau^{3}\frac{(2+\Gamma\tau)e^{-1/w^{2}}-(\Gamma t+1)}{(\Gamma\tau+1)^{2}},\label{eq:Delta_Gamma-L^2}
\end{equation}
Figure \ref{fig:FIG5}(c) shows the behavior of $\Delta_{\Gamma}L^{2}$
as a function of $w$ for various average lifetimes $\tau$. For a
given $\tau$, $\Delta_{\Gamma}L^{2}$ is positive if $w$ is sufficiently
large, i.e., 
\begin{equation}
e^{-1/w^{2}}>\frac{1+\Gamma\tau}{2+\Gamma\tau}.
\end{equation}
Note that for $\Gamma\tau\ll1$, we recover the critical width $w_{c}$
in Eq. \eqref{eq:Gaussain-w_c}. The behavior of $\Delta_{\Gamma}L^{2}$
is consistent with Fig. \ref{fig:FIG3}, where the diffusivity in
the presence of noise surpasses that in the isolated system for $w>w_{c}$.
For example, consider the curve corresponding to $w=10$ and $\Gamma=|J|$
in the right panel. In this specific scenario, the longer the wave
packet propagates in the noisy system, the further it will diffuse
compared to what would occur in a closed system.\\

\noindent \textbf{Conclusions.} In summary, our analytical study of
transient diffusivity on quantum lattices provides a wealth of predictions,
suggesting new experiments in support of the concept of the spatial
coherent control.
\begin{enumerate}
\item In comparison with a localized excitation, the finite width of the
initial state $w$ introduces spatial coherence and affects the time-dependent
diffusivity. In both closed and open systems, the difference between
diffusivity of a localized state and a finite-width wave packet scales
as $\exp(-a^{2}/w^{2})$, where $a$ is the lattice constant.
\item Gaussian or a standing wave packet with long-wavelength modulation
{[}$0\leq|k|<\pi/(4a)${]} expands slower than a delta function (width-suppressed
diffusivity), and vice versa for a small-wavelength modulation {[}$\pi/(4a)<|k|\leq\pi/(2a)${]}
(width-enhanced diffusivity). 
\item Comparing a closed and an open systems, noise can enhance transient
diffusivity of a Gaussian wave packet (including a traveling one)
of width $w>w_{c}$. The critical width $w_{c}$ is found to be of
the order of $a$ ($w_{c}\sim a$), which is satisfied under realistic
optical experimental conditions. In case of a standing wave, the critical
width depends on the wave number, and noise enhancement is possible
only for $|k|<\pi/(6a)$.
\item For a standing or traveling Gaussian wave packet exposed to environmental
noise, we establish a critical wave number/momentum value of $\pi/(4a)$,
beyond which the transient diffusivity exhibits a peak on an intermediate
time scale and exceeds the diffusivity of the initially localized
wave packet. 
\item As the traveling wave packet's initial momentum magnitude $|p|$ increases,
the center of mass motion speeds up, but the dispersion (i.e., the
diffusivity) decreases. When the momentum reaches the maximal value
$\pi/(2a)$, the noise-free wave packet becomes dispersion-less, akin
to a soliton.
\item The transient wave packet dynamics can be relevant on the time scale
of the exciton lifetime and the length scale of the diffusion length.
We explored this connection by computing the unusual dependence of
the exciton diffusion length on the initial width, dephasing rate,
and exciton lifetime.
\end{enumerate}
These theoretical predictions can be understood from two complementary
perspectives: a real space picture and a momentum space picture. (i)
The spatial coherence is negatively correlated with the transient
diffusion. A smooth wave packet (with long-wavelength modulation)
has a positive spatial correlation due to constructive interference
and thus suppresses the diffusivity. In contrast, a phase-modulated
wave packet (with a short wavelength) has a negative spatial coherence
due to destructive interference and can enhance diffusivity. The increase
in the wave packet width enhances the magnitude of the spatial coherence
and, thus, the change in the transient diffusivity relative to the
localized initial state. Furthermore, noise destroys spatial coherence
and thus suppresses the effects of spatial modulation. The noise-induced
decay of the spatial coherence results in rich and complex behavior
in the wave-packet diffusivity. (ii) We can also invoke the momentum
representation and uncertainty principle. In general, the spatial
profile of a wave packet can be transformed into a momentum distribution.
Each momentum component propagates with the group velocity determined
by the dispersion relation, and the propagation of the momentum distribution
defines the diffusivity and average velocity. This physical picture
is beneficial to understanding the diffusivity of standing and traveling
wave packets and the relationship between the two. 

Our predictions suggest various possibilities to coherently control
the wave packet dynamics on quantum lattices and have implications
for materials science and quantum technologies. Time-resolved imaging
\citep{Akselrod2014,Zhu2019,Ginsberg2020,Delor2020} of exciton dynamics
on quantum dot superlattices \citep{Blach2022,Blach2023} reveals
surprising behaviors in both the long-time and transient diffusivities,
suggesting the possibility of spatially controlling exciton wave packets.
Our analysis of the traveling wave packet can also find its relevance
in polariton transport in optical cavities \citep{EbbesenReview2021,Engelhardt2022,Cao2022,Engelhardt2023}.
While optical excitation spreads over many lattice spacings, often
exceeding the critical width of $w_{c}$, quantum technology platforms
such as optical lattices \citep{Dahan1996,Preiss2015} and super-conducting
circuits \citep{Wang2020,Gong2021,Karamlou2022} can achieve local
excitation and fine-tune the initial spreading and spatial modulation,
rendering stringent experimental tests of our predictions. These new
experiments motivate the extension of our study to disordered higher-dimensional
lattices coupled to phonons \citep{Chuang2016,Chuang2021}.

\subsection*{Associated Content}

Supporting information contains derivation details of various diffusivity
formulas and auxiliary figures.
\begin{acknowledgments}
The work is supported by the NSF (Grants No. CHE1800301 and No. CHE2324300),
and the MIT Sloan fund. We acknowledge helpful discussions with Eric
Heller, Libai Huang, Oliver Kuhn, and Keith Nelson.
\end{acknowledgments}

\bibliography{main}

\end{document}